\documentclass{osa-article}

\journal{osajournal}

\articletype{Research Article}
\begin{document}

\title{Ultrafast pulse measurement via time-domain single-pixel imaging}

\author{Jiapeng Zhao,\authormark{1,*} Jianming Dai,\authormark{2} Boris Braverman, \authormark{3} Xi-Cheng Zhang, \authormark{1} and Robert W. Boyd \authormark{1,3}}

\address{\authormark{1}The Institute of Optics, University of Rochester, Rochester, New York, 14627, USA\\
\authormark{2}Center for Terahertz Waves and School of Precision Instruments $\&$ Opto-Electronics Engineering, Tianjin University, Tianjin, 300072, PR China\\
\authormark{3}Department of Physics, University of Ottawa, Ottawa, K1N 6N5, Canada}

\email{\authormark{*}jzhao24@ur.rochester.edu} 



\begin{abstract}
In contrast with imaging using position-resolving cameras, single-pixel imaging uses a bucket detector along with spatially structured illumination for image recovery. This emerging imaging technique is a promising candidate for a broad range of applications due to high signal-to-noise ratio (SNR) and sensitivity, and applicability in a wide range of frequency bands. Here, inspired by single-pixel imaging in the spatial domain, we demonstrate a temporal single-pixel imaging (TSPI) system that covers frequency bands including both terahertz (THz) and near-infrared (NIR) region. By implementing a programmable temporal fan-out (TFO) gate based on a digital micromirror device (DMD), we can deterministically prepare temporally structured pulses with a temporal sampling size down to 16.00$\pm$0.01 fs. By inheriting the advantages in detection efficiency and sensitivity from spatial single-pixel imaging, TSPI enables the compressive recovery of a 5 fJ THz pulse and two NIR pulses with over 97$\%$ fidelity. We demonstrate that the TSPI is robust against temporal distortions in the probe pulse train as well. As a direct application, we apply TSPI to machine-learning-aided THz spectroscopy and demonstrate a high sample identification accuracy (97.5$\%$) even under low SNRs (SNR $\sim$ 10).
\end{abstract}

\section{Introduction}
{In the past decade, single-pixel imaging has emerged as a promising technique for frequency bands where high-resolution pixelated sensors are unavailable or impractical \cite{altmann2018quantum,edgar2019principles,gibson2020single}. Single-pixel imaging uses structured illumination, a bucket detector without spatial resolution, as well as computational algorithms to recover images, which is distinctive from the raster scanning-based imaging, i.e. raster scanning the transverse position of the electromagnetic wave using a bucket detector. Since multiple transverse positions are sampled in each measurement and all electromagnetic waves that have sampled the object are collected by just one single-pixel detector, a higher detection efficiency along with a lower dark count can be expected compared to imaging technique using raster scanning or cameras, which further leads to an improved sensitivity \cite{altmann2018quantum,edgar2019principles,gibson2020single}. The image acquisition time in raster scanning-based imaging scales proportionately with the number of pixels, while single-pixel imaging can efficiently sample the signal in the data acquisition process by using a computational algorithm: compressive sensing (CS) \cite{candes2008introduction}. 
CS exploits the sparsity of real-world signals by under-sampling these signals, with a number of acquired measurements that is smaller than the Nyquist-limited number of samples at the same resolution \cite{candes2008introduction}. 
Thus, CS provides the possibility of fundamentally improving the measurement efficiency and lowering the memory requirements for both data storage and transfer. }\par

\begin{figure}[htbp]
\centering
\fbox{\includegraphics[width=0.95\linewidth]{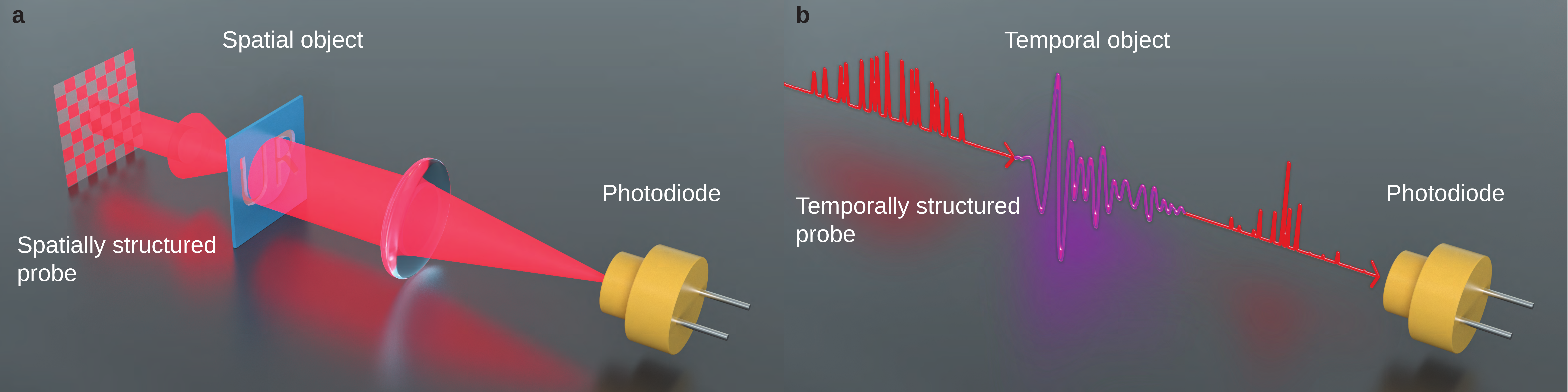}}
\caption{Comparison of single-pixel imaging and temporal single-pixel imaging. (a). A typical single-pixel imaging configuration. The photodiode is the bucket detector which has only one pixel and hence provides no spatial resolution. (b). Our proposed single-pixel temporal imaging configuration. Analogously, the slow photodiode, which lacks the temporal bandwidth to resolve ultrafast signals by itself, works as the "single-pixel" detector in the time domain.}
\label{fig:fig0}
\end{figure}

{Imaging a temporal object can be similarly accomplished through the use of a time-resolving detector. In order to temporally resolve the object, the detector should have a high bandwidth to provide a sufficient sampling rate, which is the temporal analog to the spatial resolution of a sensor array. Nevertheless, due to the limited bandwidth of electronics (usually less than 100 GHz level), imaging ultrafast signals with sub-ps-level oscillation is challenging. Numerous methods including raster scanning \cite{wu1995free}, spatio-temporal mapping on a camera \cite{teo2015invited}, frequency resolved measurements \cite{kane1993characterization, iaconis1998spectral}, spectro-temporal conversion \cite{kauffman1994time}, the use of time lenses \cite{kolner1989temporal,kolner1994space, bennett1999upconversion, salem2008optical, foster2008silicon, foster2009ultrafast, okawachi2012asynchronous,tikan2018single}, and temporal ghost imaging \cite{shirai2010temporal, ryczkowski2016ghost, devaux2016computational, o2017differential, ryczkowski2017magnified, xu2018detecting,wu2019temporal, tian2020acoustic} have been developed to bypass the limited bandwidth of electronics. Even though these methods are widely adopted in many applications, most of them can only work for selected frequency bands. For frequency bands where the devices have not been well developed, such as in the THz band, the choices of ultrafast sensing are limited compared to optical and NIR signals. The most commonly adopted THz sensing method is terahertz time-domain spectroscopy (TDS), which works as a Fourier-transform spectrometer in the THz band and relies on raster scanning \cite{wu1995free}. Even though single-shot THz sensing has been under development for years, it remains technically challenging to measure weak pulses below pJ-level due to the limited sensitivity and dynamic range of currently available cameras \cite{jiang1998single, shan2000single, kim2006single,kim2007single, kawada2011single, teo2015invited, zheng2017common}. Asynchronous optical sampling has been demonstrated to be effective for real-time THz pulse measurement as well \cite{takeshi2005asynchronous, bartels2007ultrafast}. Unfortunately, the system has to include two mode-locked ultrafast lasers with a fixed repetition frequency offset. Other techniques, for example time lenses, frequency resolved optical gating and spectral phase interferometry for direct electric-field reconstruction,
have not yet been developed for THz frequency band yet, and the current state of art of temporal ghost imaging techniques can only measure signals with ps-level oscillation, which is not accurate enough for ultrafast measurements. }\par

{Many techniques which were originally investigated in the spatial domain can inspire the development of analogous techniques in the temporal degree of freedom. The limitations in ultrafast sensing, as discussed earlier, can be resolved by using a TSPI system which is shown in Fig. 1(b). As the temporal analog of single-pixel imaging, TSPI does not place any requirements on detector bandwidth. That is to say, ultrafast electromagnetic waves can be successfully measured without any mechanical scanning using slow detectors, which are "single-pixel" detectors in the time domain working at corresponding frequency bands. Furthermore, TSPI inherits other advantages of spatial single-pixel imaging: increased sampling efficiency, flexibility for diverse wavelengths and high sensitivity for weak signals. Although TSPI can link the ultrafast optical signals with slow electronics, previous demonstrations did not achieve sub-ps-level ultrafast sensing due to the difficulty of deterministically preparing temporally structured pulses with fs-level modulation \cite{xu2018detecting}. Here, we develop a programmable temporal fan-out (TFO) gate based on a commercially available DMD, with a minimum temporal sampling size (the temporal interval between two adjacent TFO replicas) $\Delta \tau = 16.00\pm$0.01 fs. As the temporal analog of spatial fan-out, which can duplicate an incoming light pulse into spatially separated replicas \cite{dammann1971high,prongue1992optimized, romero2007theory, mirhosseini2013efficient}, TFO duplicates the input ultrafast pulse into coherent replicas separated in the time domain, forming an ultrafast pulse train. Since the TFO gate is programmable, an ultrafast pulse train with a variable temporal interval and arbitrary temporal structure can be deterministically prepared. It is worth noting that our convenient temporal device not only allows TSPI to image ultrafast signals, but also provides new possibilities in ultrafast waveform synthesis, as well as atomic and molecular control. }\par

\section{Results}
\subsection{Experimental Configuration}

\begin{figure}[htbp]
\centering
\fbox{\includegraphics[width=0.95\linewidth]{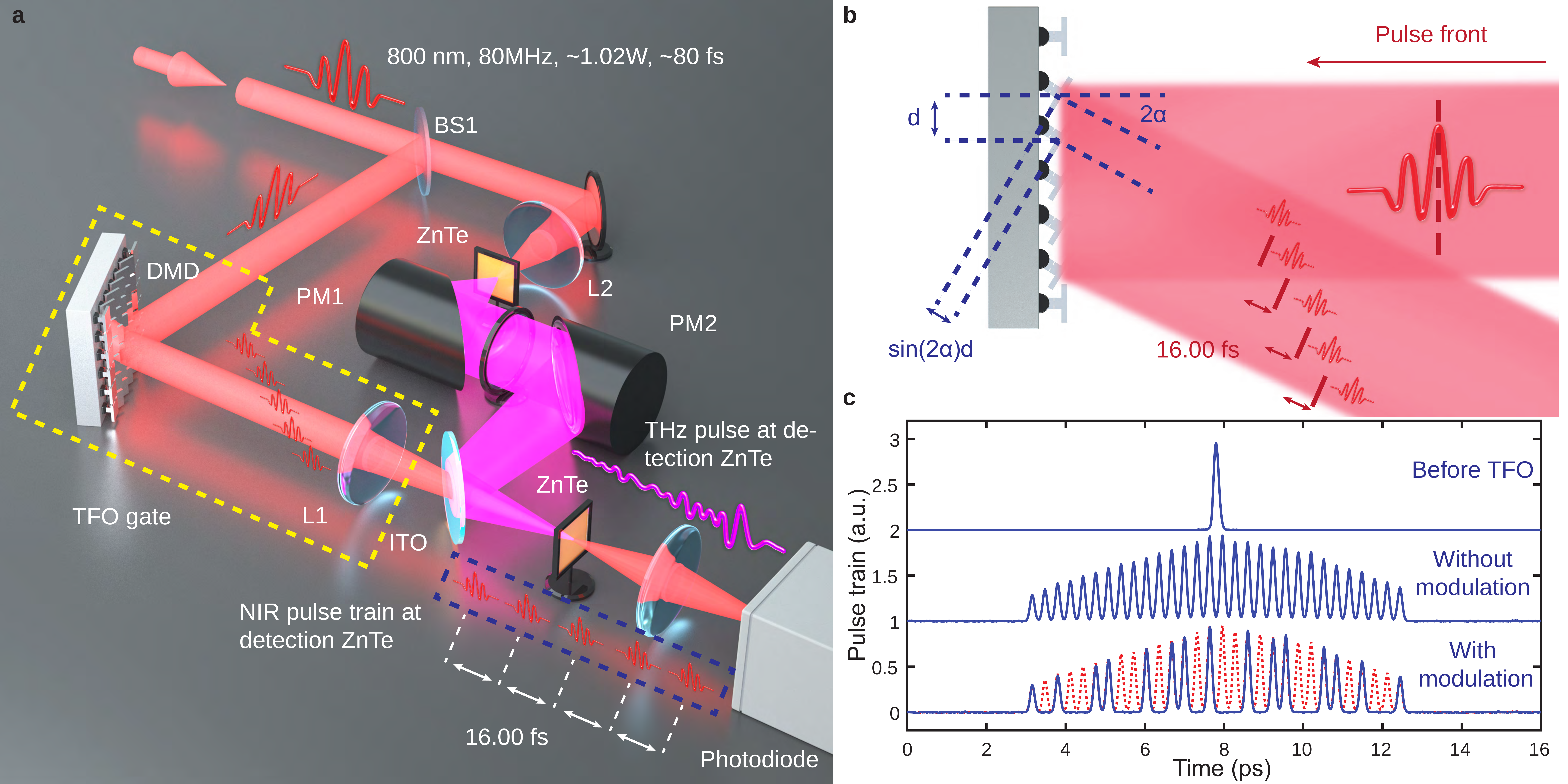}}
\caption{Schematics of the experimental setup and the structure of the TFO gate. (a): Experimental configuration. The yellow dashed box indicates our TFO gate, and the blue dashed box indicates the prepared ultrafast pulse train at the detection ZnTe crystal plane. (b): A detailed sketch illustrates the layout of DMD and how its geometry leads to our TFO. The dark red arrow shows the propagation direction of input ultrafast pulse. $\alpha = 0.21$ rad is the tilt angle of each micromirror while $d = 10.8$ $\mu$m is the separation between two micromirrors. Dark red lines represent the corresponding pulse front. The $\Delta \tau$ between two TFO copies, is given by: $\Delta \tau = \text{sin}(2\alpha)d/c = 14.66$ fs, where $c$ is the speed of light in air. Note that, in the experiment, the light is normally incident on the DMD. (c): Examples showing how the TFO works. Each effective TFO replica consists of 5 DMD columns. Due to the long pulse duration of the NIR pulse, the separation between two effective replicas is 240 fs (15 DMD columns) to explicitly show the structure of pulse trains. The red dashed trace indicates that the TFO copies at those temporal position are turned off by modulating the TFO gate. All cross-correlation traces of pulse trains are averaged over 9 measurements.}
\label{fig:fig1}
\end{figure}

Fig. 2(a) schematically illustrates the concept of our TSPI scheme. One terahertz (THz) pulse with about 3 V/cm peak electric field and 5 fJ pulse energy works as the temporal object in the current system. Our TFO gate is shown in the yellow dashed box, which consists of a DMD (Texas Instruments DLP3000) and a lens (L1). The geometry and layout of the DMD are shown in Fig. 2(b). Before the DMD, there is no time delay in the pulse front of the incident pulse. After the DMD, $N$ spatio-temporally separated fan-out replicas are generated \cite{murate2018adaptive}. L1 is used to perform a Fourier transform so that, in the Fourier plane, the input ultrafast pulse is converted to pulse trains containing $N$ TFO copies at the same transverse position without any spatial-temporally coupled phase from diffraction. $N$ depends on the number of DMD columns illuminated by the incident pulse, which is 608 in the current system. The intensity of each replica can be arbitrarily tuned by enabling only a subset of the mirrors in each DMD column, which further enables arbitrary control over the pulse train (examples are shown in Fig. S2). $\Delta \tau$ is determined by the physical size and tilt angle of the micromirrors. For the current device, the measured $\Delta \tau$ is 16.00$\pm$0.01 fs, which is consistent with the theoretical value of 14.66 fs found in the caption of Fig. 2. Therefore, the upper bound of the Nyquist frequency of our TSPI is 31.25 THz, which is sufficient for measuring most broadband pulses. A larger $\Delta \tau$ and a longer pulse duration of each TFO replica can be achieved by combining multiple DMD columns into a single "effective column" due to the interference between adjacent TFO replicas. The total time window $T$ {(the temporal field-of-view)} depends on the overall size of the mirror array, and is equal to $N \times \Delta \tau$, which is 9.73 ps. It is noteworthy that, by selecting a DMD chip with the appropriate specifications, different $\Delta \tau$, $N$ and $T$ can be achieved as well. The encoding speed is mainly limited by the switching speed of the DMD, which can be as large as 4 kHz in our system and can be further increased to 20 kHz by using a faster DMD chip \cite{stantchev2020real}. One of the beauties of TSPI is that the temporal sampling size comes from the modulation on TFO but not from the bandwidth of detectors. Therefore, our scheme solely requires detectors with bandwidth exceeding the modulation frequency of the DMD, allowing the use of kHz-level photodiodes, which have better noise-equivalent power than GHz-level fast photodiodes. {As a comparison, previous temporal imaging schemes, including the use of time lenses and temporal ghost imaging, cannot easily reach such high effective sampling rate and sensitivity even with the use of fast detectors with bandwidths at 100 GHz level \cite{kolner1989temporal,kolner1994space, bennett1999upconversion, salem2008optical, foster2008silicon, foster2009ultrafast, okawachi2012asynchronous, shirai2010temporal, ryczkowski2016ghost, devaux2016computational, o2017differential, ryczkowski2017magnified, xu2018detecting,tikan2018single, wu2019temporal,tian2020acoustic}.} Thanks to encoding the probe beam as opposed to temporally modulating the THz pulses \cite{zhao2019spatial}, our simple and reliable scheme can directly utilize commercially available DMDs with a high modulation speed and damage threshold, and hence no additional fabrication of devices is required. \par

\begin{figure}[htbp]
\centering
\fbox{\includegraphics[width=0.95\linewidth]{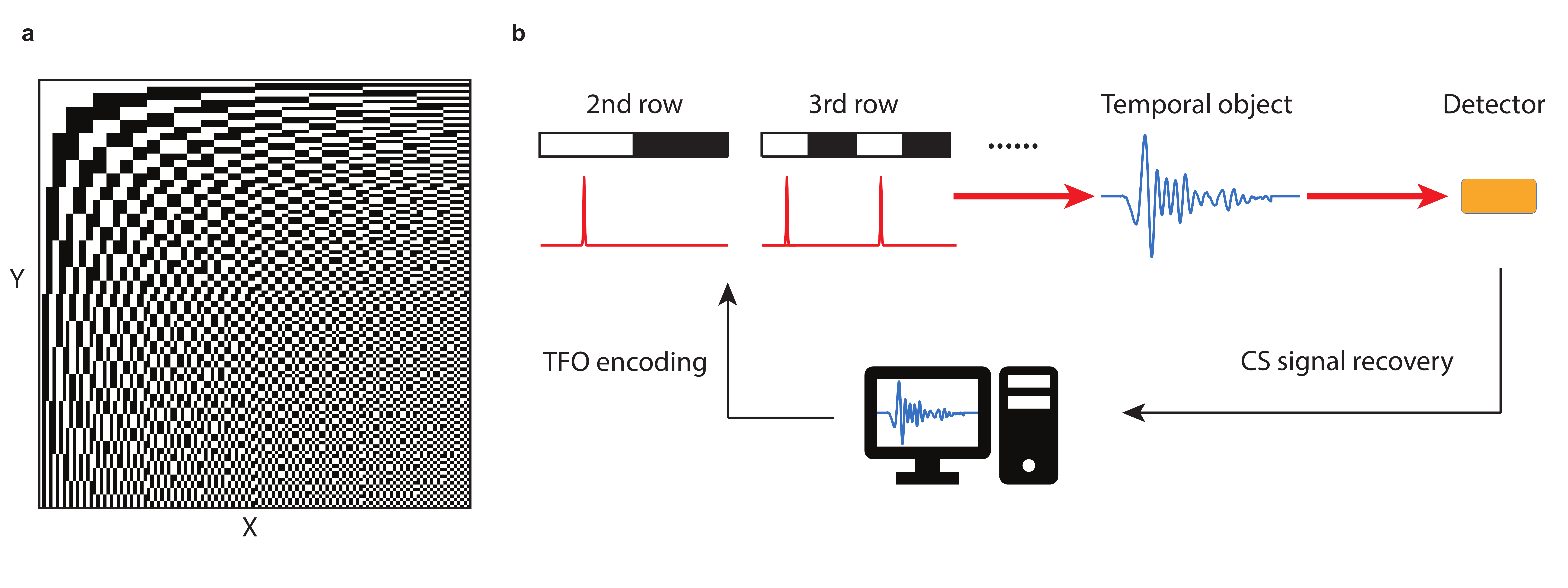}}
\caption{Walsh-ordered Hadamard matrix and the flow-chart of the sampling and data processing based on CS. (a): A 128-dimensional Walsh-ordered Hadamard matrix, which is used to sample the temporal object in our experiment. (b): The flow-chart of the experiment. The probe pulse is temporally modulated by TFO based on the row vectors of the Walsh-ordered Hadamard matrix. }
\label{fig:fig2}
\end{figure}

\subsection{Compressive sensing in ultrafast measurements}

Before we briefly introduce the concept of compressive sensing (CS), a computational algorithm for accurate reconstruction of sparse signals, it is convenient to revisit the measurement process in the conventional ultrafast sensing. If we assume that our pulse $E(t)$ is temporally sampled at $M$ time points, the analog signal $E(t)$ will be represented by a $M$ dimensional, digitized vector $E_d(t) = [E(t_1),E(t_2),...,E(t_M)]$. This $M$ dimensional vector $E_d(t)$ is sampled by a sampling matrix $S$, which is the $M$ dimensional identity matrix in the case of traditional ultrafast sensing techniques based on either raster scanning or high-bandwidth detectors. This identity sampling matrix corresponds to the "natural basis", where the $i$th sampling row vector represents the sampling at temporal position $t_i$, and the sampling is experimentally accomplished by using a high-bandwidth detector to measure the pulse at $t_i$, or raster scanning a slow detector to measure the pulse at $t_i$. Mathematically, the measurement process can be represented by: $\phi = SE_d(t)$, where $\phi$ is the coefficient vector corresponding to the data recorded by the detector in the experiment. Thus, recovering the digitized pulse $E_d(t)$ based on experimental data $\phi$ and sampling matrix $S$ is simply given by: $E_d(t) = S^{-1} \phi$. Since the sampling basis $S$ is the identity matrix in the conventional ultrafast sensing, the experimental measurement result $\phi$ directly represents the digitized pulse $E_d(t)$.\par

From the discussion above, it is obvious that at least $M$ measurements, which is the minimal number of measurements required by Nyquist sampling limit, are required to recover the signal because the sampling matrix $S$ has to be inverted. However, by adopting CS, we can significantly reduce the number of measurements to a number which is smaller than $M$. Since most signals in nature are sparse when represented in an appropriate basis \cite{candes2008introduction}, CS can take advantage of this sparsity and find the proper sparse basis $R$ so that most coefficients in $RE_d(t)$ are close to zero \cite{candes2008introduction}. Therefore, one can recover most information of the signal vector $E_d(t)$ by only measuring those nonzero coefficients in $RE_d(t)$. This indicates that in the experiment, we only need to use sampling vectors whose corresponding coefficients are nonzero to sample the unknown ultrafast signal. Hence, a number of measurements that is less than $M$ is sufficient to recover the signal without losing too much information, which surpasses the Nyquist sampling limit because of this under-sampling measurement procedure \cite{candes2008introduction}.\par

In order to fulfill the condition of CS, $R$ has to be incoherent to basis $S$, which means that the correlation between any two elements in $R$ and $S$ has to be small. Luckily, if $R$ is the Fourier conjugate basis of $S$, they are always maximally incoherent to each other (a detailed discussion of incoherent basis and choice of sampling basis can be found in Supplementary Section 1) \cite{candes2008introduction}. Furthermore, the majority of information in THz signals is carried by low frequency terms. 
Based on these two facts, if we use $S$ as the natural basis representing the sampling of temporal position, we will use Walsh-ordered Hadamard matrix as the sparse sampling basis $R$ in our experiment. This is because that the Walsh-ordered Hadamard transform is a generalized discrete Fourier transform, and hence the row vectors of the Walsh-ordered Hadamard matrix represent different frequencies in an increasing order. Therefore, we can either sequentially use each row of the Walsh-ordered Hadamard matrix to sample the ultrafast signal from the low frequency to high frequency components, or selectively measure one or more arbitrary frequencies by choosing the corresponing row vectors to encode the probe pulse train. Here, we mainly show the result of sequential sampling. \par 

The procedure of the sampling and data processing is shown in Fig. 3(b). We first use each row vector of the 128-dimensional Walsh-ordered Hadamard matrix (shown in Fig. 3(a)) to modulate the pulse train via TFO. The encoded pulse train is then used to sample the temporal object, which is a 5 fJ THz pulse. After the interaction between two pulses, the signal is recorded by a slow photodiode, which works as the "single-pixel" detector in the time domain. By correlating the rows of the sampling matrix and the corresponding signal recorded on the detector, we can recover the temporal object.\par

\begin{figure}[htbp]
\centering
\fbox{\includegraphics[width=0.95\linewidth]{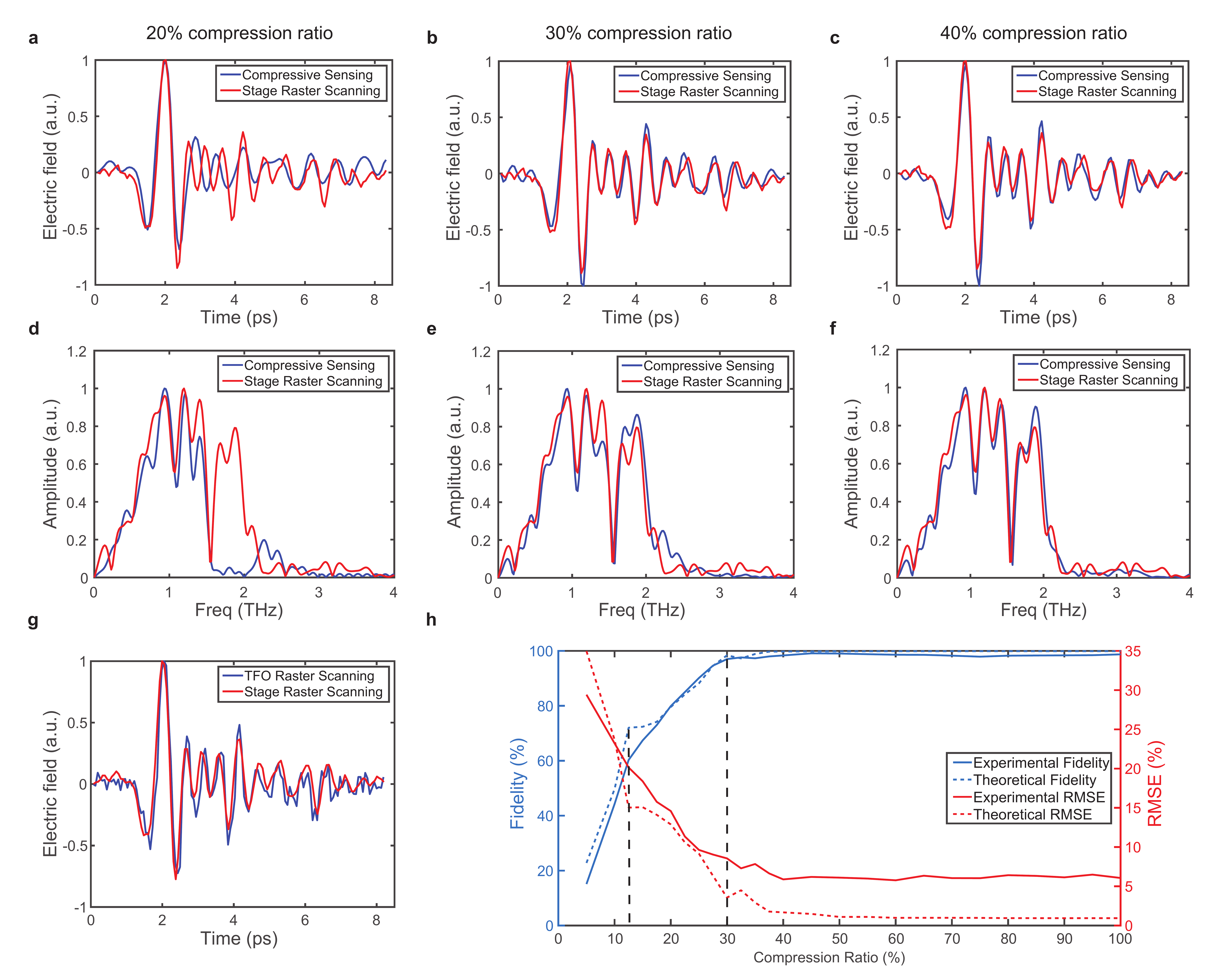}}
\caption{Recovered THz electric fields and spectra at different CRs (blue curves). The THz fields and spectra measured by raster scanning a mechanical delay stage (red curves) are shown for comparison. (a) and (d): Recovered THz field and spectrum at 20$\%$ CR. The fidelity in THz field is 84.30$\%$ and the root mean square error (RMSE) is 14.92$\%$. (b) and (e): Recovered THz field and spectrum at 30$\%$ CR. The fidelity in THz field is 97.43$\%$ and the RMSE is 7.69$\%$. (c) and (f): Recovered THz field and spectrum at 40$\%$ CR. The fidelity in THz field is 97.76$\%$ and the RMSE is 7.17$\%$. THz pulses recovered by both CS and raster scanning are measured without averaging and use the same $\Delta \tau$ (64 fs). Due to the limited detection bandwidth of ZnTe crystal, we only show the spectra in 0-4 THz range. (g): Raster scanning using TFO. The sampling rate and acquisition time of each measurement are set to be the same as CS data. Fidelity is 87.35$\%$ and RMSE is 13.11$\%$. (h): The measured and theoretical fidelities and RMSEs as functions of CR. The RMSE is mainly limited by the measurement noise.}
\label{fig:fig2}
\end{figure}

\subsection{Experimental results}
Fig. 4 shows the THz pulses reconstructed based on TSPI and the flow-chart in Fig. 3. The quantity describing the level of compression is termed as compression ratio (CR), which is defined as the ratio between the number of measurements taken in the experiment over $M$, i.e. measuring all the rows of the Hadamard matrix. Limited by the pulse duration of the original pulse (80 fs), we choose $\Delta \tau$ to be 64 fs by combining 4 DMD columns into 1 effective column. 128 TFO replicas, encoded by the 128-dimensional Walsh-ordered Hadamard matrix, constitute our probing pulse train with $T=8.19$ ps. The red curves are THz fields and spectra measured by raster scanning a mechanical delay stage without averaging. As illustrated in Fig. 4(d), by sacrificing high frequency components, a recovered THz pulse with high fidelity (84.30$\%$) can be achieved even at 20$\%$ CR. {Fidelity is defined as the correlation between the THz fields (spectra) measured by raster scanning and the THz fields (spectra) recovered by CS.} At this CR of 20$\%$, the last sampling pulse train is encoded by the 25th row of the Walsh-ordered Hadamard matrix, which results in the cutoff sampling frequency at 1.5 THz and a weak sidelobe at 2.4 THz. For a CR of 30$\%$, additional high-frequency terms are measured and in the spectrum, almost all high-frequency terms in THz pulse are collected, leading to a near-unity fidelity of 97.43$\%$ in THz field. As the CR goes higher to 40$\%$, more high-frequency components are collected but the improvement is limited (fidelity of 97.76$\%$ in field). When we look at the spectrum, two spectra match well for all frequencies except for some noise, which indicates that we have included enough high frequencies in the measurement. Therefore, all higher-frequency terms can be ignored due to their relatively small contributions. \par

{Comparing to the data acquisition time of the mechanical stage raster scanning data {(sampling rate at 2 Hz which is mainly limited by the stability of our mechanical stage, leading to a measurement time of $\sim$64 s)}, our TSPI system requires only $\sim$15 s with 30$\%$ CR {(sampling rate at 10 Hz)}, and can be further reduced below 1 s by upgrading the experimental hardware (See Supplementary Section 6). Thanks to the fewer measurements enabled by CS and the faster sampling speed provided by DMD, TSPI can reduce data acquisition time, data storage and transfer memory requirements by a factor of at least 2/3, which would significantly enhance system efficiency in this era of big data. {It is also worth noting that the data processing time is $\sim$ 1 s for a 30$\%$ CR result using MATLAB on a normal laptop, which will become longer when the CR is higher (the upper limit is less than 2 s). A shorter processing time should be available when the MATLAB code is further optimized and executed on a more powerful machine.}
Apart from a more efficient measurement, CS is also more robust to noise than raster scanning. As shown in Fig. 4(g), under the same sampling rate and integration time of each measurement, the fidelity of the TFO raster scanning data, i.e. sequentially turning on each DMD column, is only 87.35$\%$ and the RMSE is 13.11$\%$, which are worse than the CS data even when the CR is only 30$\%$.}\par

The experimental results of the fidelity and RMSE as a function of CR, shown in Fig. 4(h), are in good agreement with our theoretical predictions, which is shown in Supplementary Fig. S1. As the CR increases, both fidelity and RMSE change dramatically at first, and then gradually level out after the CR reaches 30$\%$. Our theoretical model also predicts that there are two discontinuity points in the slope of both RMSE and fidelity curves. As indicated by the two vertical dashed lines, the first turning point is located at CR equal to 12.5$\%$ while the second one is at CR equal to 30$\%$. The slopes in both RMSE and fidelity curves abruptly decrease to a slower rate and can even change sign after the turning points. Similar observations have been found in spatial single-pixel imaging work when the sampling matrix is reordered \cite{sun2017russian}. Therefore, in both spatial and temporal CS, one can optimize the acquired information within a given CR by rearranging the sampling order, which demonstrates the spatio-temporal duality of light pulses. An improved measurement efficiency can be predicted by further optimizing the sampling strategy based on these turning points.\par 

\subsection{Robustness Against Temporal Distortions}

\begin{figure}[htbp]
\centering
\fbox{\includegraphics[width=0.95\linewidth]{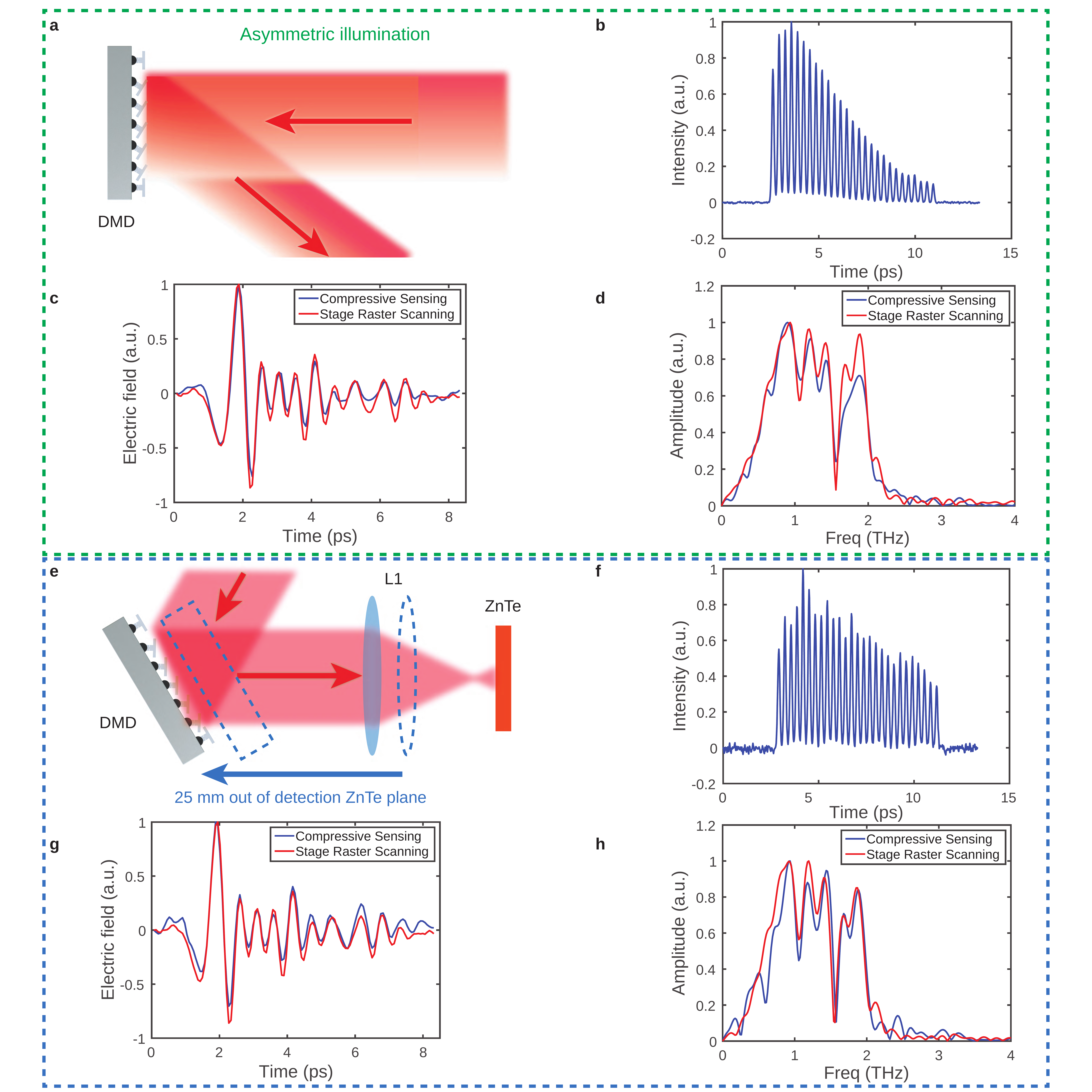}}
\caption{Recovered THz electric fields and spectra using distorted TFO gate at 40$\%$ CR. Red arrows illustrate the propagation direction of NIR pulses. (a): How we distort TFO gate with an asymmetric illumination. (b): The measured asymmetric pulse train formed by the distorted TFO copies. The ratio between the maximal and minimal intensities in (b) is about 10. As a comparison, this ratio is about 3 in Fig. 2(c) while the SNR is the same ($\sim$350). (c) and (d): The corresponding recovered THz field and spectrum. The RMSE in field is 7.21$\%$. (e): How we move the Fourier plane of DMD 25 mm away from the detection ZnTe plane by moving TFO gate in the direction as the blue arrow shows. (f): The measured pulse train when the detection ZnTe crystal is 25 mm out of the Fourier plane of DMD. The intensity envelope becomes irregular and SNR is poor ($\sim$80). (g) and (h): Recovered THz pulse and spectrum using the distorted pulse train in (f). The RMSE in field is 7.66$\%$. (b) and (f) are averaged results of 9 measurements while THz pulses recovered by CS and raster scanning have no averaging. All results shown above are consistent with theoretical predictions in Supplementary Section 4. }
\label{fig:fig3}
\end{figure}

Similar to its spatial counterpart, TSPI is resistant to temporal distortions. To demonstrate this robustness, we first intentionally make the illumination on the DMD asymmetric, which is shown in Fig. 5(a). This results in a different intensity for each TFO replica, and further leads to asymmetric pulse trains in the time domain (Fig. 5(b)). Despite the distorted TFO replicas, the recovered THz signal, shown in Fig. 5(c) and (d), has a fidelity of 96.96$\%$ in the field at 40$\%$ CR. The major information loss comes from underestimation of oscillations at the tail of the THz pulse (above 5 ps), which results from the weak intensities of TFO replicas at the end of the pulse train. Next, as illustrated by the blue arrow in Fig. 5(e), we move the location of the Fourier plane of DMD 25 mm away from the ZnTe crystal by moving the TFO gate. Due to the spread of k-vectors of each replica at the Fourier plane, the TFO copies will be spatio-temporally separated at the detection crystal plane. This leads to the spatio-temporal coupled sampling, which yields an irregular pulse train envelope and lowers the intensity of each sub-pulse (Fig. 5(f)). Nevertheless, as shown in Fig. 5(g) and (h), distorted pulse trains do not affect the quality of the recovered signal, which has a fidelity of 98.34$\%$ in field at 40$\%$ CR. Such a robustness can also be interpreted by the spatio-temporal duality of light pulses. In single-pixel imaging, the spatial correlation between the intensity distribution and position lies at the heart of the successful recovery of images. {A distortion in the intensity profile of the illumination distribution will lead to the over- or under-estimation of the object intensity, but the accuracy of signal recovery will not decay significantly if CS is used. Similarly, as long as the correlation between temporal position of each TFO replica and intensity envelope of the entire pulse train is not destroyed, the THz pulse recovered by TSPI will be robust against temporal distortions, which is numerically demonstrated in the Supplementary Section 4.}\par

\subsection{NIR pulses with different pulse durations}

\begin{figure}[htbp]
\centering
\fbox{\includegraphics[width=0.95\linewidth]{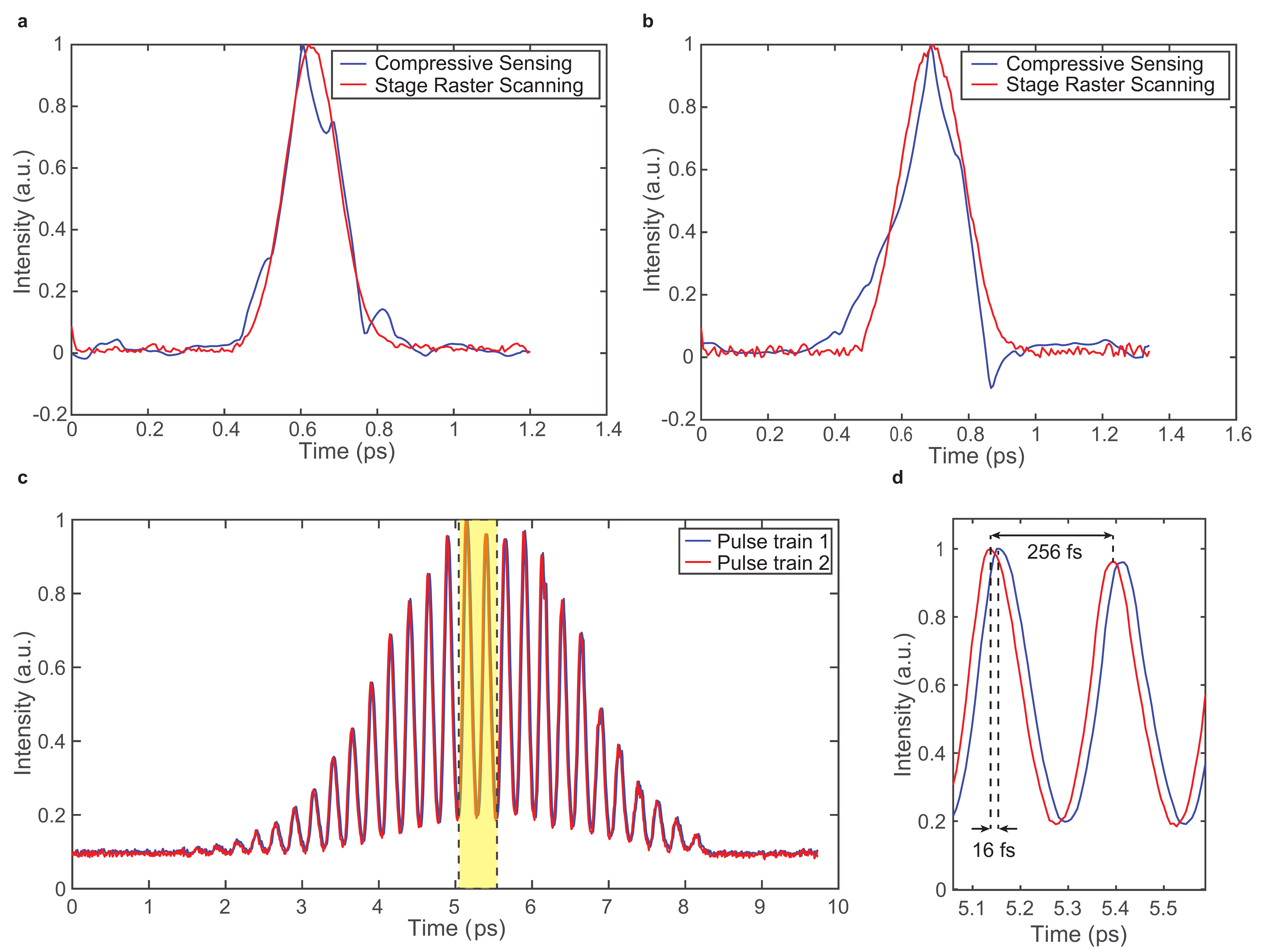}}
\caption{Recovered NIR pulses (blue curves) and the measured results using raster scanning (red curves). (a): Recovered NIR pulse with 90 fs pulse duration at 80$\%$ CR. (b): Recovered NIR pulse with 125 fs pulse duration at 80$\%$ CR. (c): Two pulse trains used for NIR measurement. By shifting 1 DMD column, we shift pulse train 1 16 fs forward to get pulse train 2. (d): The details of the measured pulse trains shown in the black dashed box in (c). One can find that the displacement between two pulse trains is 16 fs while the separation between two peaks in one pulse train is 256 fs. }
\label{fig:fig6}
\end{figure}

To demonstrate that TSPI can work for signals in a broad range of frequencies, we recover two NIR pulses with $\sim$90 fs and $\sim$125 fs pulse durations. To perform the measurement, the detection crystal is changed to a 1-mm-thick BBO crystal. A photomultiplier tube is used as the new detector to replace the balanced photodiodes in THz measurements. Due to the long pulse duration of our laser (80 fs), preparing high-contrast sampling pulse trains with 16 fs temporal modulation size is not feasible. Therefore, we have to implement a rolling-average strategy to measure NIR pulses with 16 fs sampling size. In order to do this, we first combine 5 DMD columns together (80.00 fs) as an effective column. Then we perform the measurement of the unknown pulse in one time window. Next, we shift this time window 16.00 fs forward to make another set of measurement, which corresponds to the shift of 1 DMD column. We continue this rolling process until 5 shifts are done. The final recovered NIR pulse is the rolling average of the measurements of all 5 shifts.\par

The experimental results are shown in Fig. 6. We change the pulse duration in the signal arm by pre-chirping the laser pulse using the self-phase modulation effect inside a ZnTe crystal. In the meantime, the pulse duration of the probe arm is kept at 80 fs. Fig. 6(a) is the compressive measurement result (blue curve) of a 90 fs pulse comparing to the delay stage result (red curve). The recovered signal has a fidelity of 98.48$\%$ in pulse shape while the pulse duration is 91.4 fs. Fig. 6(b) shows the compressive result of a 125 fs pulse. The fidelity is 97.88$\%$ while the pulse duration is 110.0 fs. Both CS results are measured at 80$\%$ CR due to the fact that NIR pulses carries more high-frequency components. The main limiting factor in NIR pulse measurement is the instability of the signal pulse induced by self-phase modulation. This instability not only leads to a worse SNR (less than 70 in delay stage data), but also results in a irregular but large intensity fluctuations at the peak. When these significant sources of noise are included in CS recovery, we will have a distorted pulse. For example, in the imperfect reconstruction of the 125 fs pulse, we have observed a negative sidelobe at 0.85 ps in Fig. 6(b), which has no physical meaning because the measured quantity is the pulse intensity. We expect that this noise should vanish if one uses linear optics devices to introduce chirp. Therefore, a better recovery should be available for real NIR pulses whose pulse duration are not extended by nonlinear effects. To get a more accurate CS recovery, one can also intentionally stretch the NIR pulses, for example using time lenses, to a longer level and then use our TSPI to measure it \cite{kolner1989temporal,kolner1994space, bennett1999upconversion, salem2008optical, foster2008silicon, foster2009ultrafast, okawachi2012asynchronous,tikan2018single}. It is notable that the TSPI system can be used for other wavelengths as well. One simply needs to replace the crystal and detector with the ones at the correct frequency band. \par

\subsection{Machine-Learning-Aided THz Spectroscopy}
\begin{figure}[htbp]
\centering
\fbox{\includegraphics[width=0.95\linewidth]{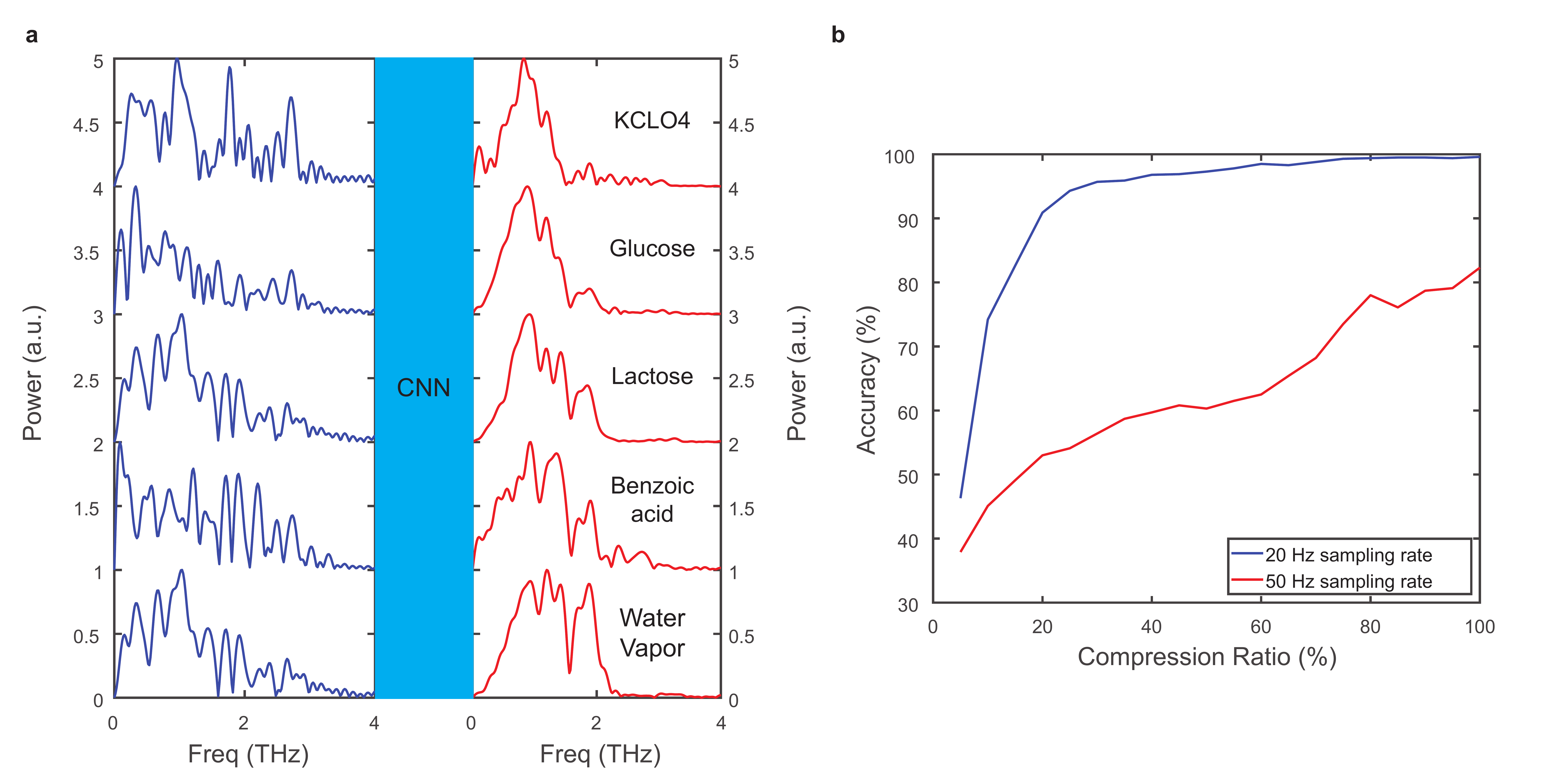}}
\caption{CNN-enhanced THz spectroscopy. (a): The sample classification is based on a CNN. Due to the poor SNR, the fingerprints of each sample are lost and cannot be identified from spectra (blue curves), which leads to a classification accuracy of 34.2$\%$ without the use of ML. Red curves are desired spectra that we are trying to match. After training the CNN, we have an average classification accuracy of 96.8$\%$ at 40$\%$ CR. (b): Accuracy as a functions of CR. When the sampling rate is 20 Hz, the accuracy rises up quickly when the CR increases from 5$\%$ to 20$\%$, and then starts to converge, which has a very similar behavior as the fidelity shown in Fig. 4(h). As a comparison, when sampling rate is 50 Hz, the SNR in each measurement becomes 1.73 times smaller than the 20 Hz sampling rate case, and accuracy curve does not start converging even when CR is 100$\%$. All the identification accuracies are averaged over 5 CNN trainings. }
\label{fig:fig7}
\end{figure}

Lastly, we demonstrate a compressive ultrafast THz TDS system as one potential application of TSPI, and show that we can improve its performance through use of machine learning (ML). Five different samples (water vapor, glucose, lactose, benzoic acid and $\text{KCLO}_4$) are tested in the system, and 1000 sets of data of each sample are used for training. To show the noise resistance of the ML-enhanced TDS system, the SNR of the THz pulse is decreased to $\sim$10 by attenuating the THz pulse to sub-fJ level and lowering the integration time on the lock-in amplifier. As shown in Fig. 7(a), without the help of ML, the measured THz spectra are noisy, and by calculating the maximum fidelity to the reference waveforms, the classification accuracy is only 34.2$\%$. As a comparison, after training our convolutional neural network (CNN) with 5000 sets of data, we can successfully classify the remaining 2000 sets of data (400 for each sample) with an accuracy of 97.50$\%$ at 40$\%$ CR (see Supplementary Section 5 for confusion matrices). {If one further reduces the SNR by increasing the sampling rate from 20 Hz to 50 Hz, the identification accuracy will significantly decrease as shown in Fig. 7(b). Therefore, if training data sets are measured using other approaches with worse robustness against noise, for instance raster scanning, a high identification accuracy will not be achieved even with the help of ML.}  \par

\section{Discussion and conclusion}
Raster scanning the time delay between a short probe pulse and the signal pulse is the conventional approach for ultrafast sensing. As a contrast, TSPI can significantly reduce the data acquisition time in the sampling process, and provide a better fidelity as well as an enhanced robustness against noise (the comparison in Fig. 4(g)). \par

Comparing to methods using cameras or spectrometers, TSPI is applicable to both weak and strong pulses at various frequency bands. Even though cameras and spectrometers enable the possibility of single-shot measurements, effective measurement of weak pulses is not applicable at some wavebands. For example, in THz region, these single-shot measurement methods usually require the use of $\mu$J-to-mJ-level pump pulse to generate pJ-to-nJ-level THz pulses \cite{jiang1998single, shan2000single, kim2006single,kim2007single, kawada2011single, teo2015invited, zheng2017common}. As a comparison, our approach only needs nJ-level pump pulses, and is capable of measuring fJ-level THz pulses. Hence, the sensitivity of our system is about 3 orders of magnitude better than the camera- or spectrometer-based methods. Furthermore, when pulses become weak, these single-shot measurement methods will need to average over multiple shots to provide a good SNR, which is no longer a single-shot measurement and will result in a slow measurement speed as well. However, for NIR and optical pulse measurements, our method does not presently resolve spectral phase and minimal resolvable pulse duration cannot be below 10 fs. Therefore, even our approach is applicable for weak THz pulses, the current demonstration cannot provide comprehensive characterization of NIR and optical pulses compared to frequency-resolved optical gating and spectral phase interferometry for direct electric-field reconstruction \cite{kane1993characterization,iaconis1998spectral}. The possible spectral phase measurement might be available by introducing another phase-only SLM to the probe arm to include phase encoding. \par

Temporal imaging based on time lenses is another way to measure ultrafast signals. By stretching the pulse to a longer duration, one can use fast photodiodes to measure the optical pulses \cite{kolner1989temporal,kolner1994space, bennett1999upconversion, salem2008optical, foster2008silicon, foster2009ultrafast, okawachi2012asynchronous,tikan2018single}. Even though the ultrafast measurement done by time lenses can be much faster than our computational-based TSPI, our technique is still advantageous in terms of the working frequency bands and flexibility. The working frequency bands of time lenses is usually at optical and NIR regions, and time lenses in the THz band have not been demonstrated yet. Meanwhile, even in the optical and NIR wavebands, the design and fabrication of photonic time lenses have to be changed for each wavelength, which usually lacks flexibility compared to our TSPI.  \par

It is worth noting that a concise THz single-pixel hyperspectral imaging system can be available by combining TSPI and our previous spatial sampling system \cite{zhao2019spatial}. Since both approaches are based on the probe-beam encoding, one can use multiple DMDs in the probe arm to deterministically encode spatial and temporal structures onto the NIR probe beam. By implementing CS and such a spatio-temporally structured probe beam, it is possible to comprehensively and compressively measure the THz pulse with a high measurement efficiency, which can significantly reduce the data acquisition time and data storage requirement in conventional hyperspectral imaging systems.\par

In summary, based on an ultrafast TFO gate, we develop a TSPI system to compressively measure ultrafast signals using exclusively kHz-rate slow detectors. The 16.00$\pm$0.01 fs temporal sampling size enables the successful measurement of a 5 fJ THz pulse and two NIR pulses with high fidelity. The robustness of TSPI against temporal distortion is also demonstrated. Lastly, we show that our technique can be used to perform ultrafast THz TDS with the help of ML. Our TSPI system can cover a frequency band including both NIR and THz region with a high sensitivity, efficiency and robustness. This distinctive technique can provide dynamic temporal imaging of ultrafast signals, leading to potential applications including single-pixel hyperspectral imaging, remote sensing, and high-data-rate optical communications.\par

\section{Methods}

Here we include additional details of our experimental setup. As shown in Fig. 2(a), a 800 nm NIR laser pulse (Maitai oscillator laser from Spectra-Physics) with 80 fs pulse duration and 80 MHz repetition rate is split by a 45:55 non-polarizing beam splitter (BS1). The average power is 1.02 W which corresponds to 12.75 nJ pulse energy. The diameter of the probe pulse is expanded to 30 mm by two lenses with focal lengths equal to 75 mm and 500 mm, respectively (not shown in figure). One rectangular aperture stop (3.7 mm by 6.6 mm, not shown in figure) is used to select the central part of the expanded probe pulse, providing a relatively uniform intensity distribution. The selected probe pulse is then incident on the DMD (Texas Instruments DLP3000). Note that, our DMD chip is the most basic one which has a medium micromirror size, a small amount of micromirrors and a small total size of the mirror array. Therefore, by using a more advanced commercially available DMD with a smaller mirror size (a shorter $\Delta \tau$, e.g. $\sim$11 fs), a larger $N$ (e.g. 2560) and a larger temporal field-of-view (a larger $T$, e.g. $\sim$28 ps), one can easily extend the potential applications of our scheme. After DMD, a 250-mm-focal-length lens (L1) is used to perform the Fourier transform, and the detection ZnTe crystal (10$\times$10$\times$3 mm) is placed at the Fourier plane. {Although the preparation of pulse trains comes from the focusing in the horizontal direction, the quadratic phase from the diffraction can only be eliminated if the Fourier transform is performed.} In the signal arm, a delay stage (not shown in the figure) with two mirrors is used to measure THz waveforms (red curves in Fig. 4 and 5) for comparison. The NIR pulse is then focused onto an 1-mm-thick ZnTe crystal using a 75-mm focal length lens (L2) to generate THz pulse. Two 4-inch-focal-length parabolic mirrors (PM1 and PM2) are used to collimate and refocus THz pulses onto the detection ZnTe. An ITO-coated glass plate is used as a dichroic mirror to combine THz and probe beams. After the detection crystal, a conventional electro-optic sampling detection unit is used, which consists of a quarter-wave plate, a Wollaston prism and a balanced photodiode detector (Hamamatsu Photonics S2386-8K). The signal from the balanced photodiode detector is finally measured by a lock-in amplifier (Stanford Research Systems SR830). The acquisition time of each measurement (one pattern on DMD or one temporal position on delay stage) in either TSPI or delay stage raster scanning is the same (100 ms). However, the integration time (time constant) on lock-in amplifier is different. The delay stage raster scanning has a integration time of 100 ms while the TSPI data is 30 ms to avoid the rolling average on lock-in amplifier. Therefore, the SNR of each measurement in delay stage raster scanning is 1.83 times higher than the TSPI data. All experimental raw data and numerical codes are available upon reasonable request. \par

Compared to the minimal $\Delta \tau$ of the pulse train, our laser pulse has a relatively long pulse duration of 80 fs. {Therefore, adjacent replicas will be highly overlapped in time. Due to the fact that TFO replicas are mutually coherent, this overlap will lead to multi-pulse interference and the adjacent copies will coherently add up to form a new pulse. Thus, we can combine every 4 DMD columns together as effective columns to prepare pulse trains with a $\Delta \tau=64$ fs.} For the same reason, to avoid the multi-pulse interference between adjacent TFO replicas, we represent each row of Hadamard matrix $H$ as the sum of four sub-rows. Letting the Nth row be defined as $\{$$a_i$$\}$, we set the sub-rows as $\{$ $b_i=a_i$, for i= 0,1,2,3 mod 4 $\}$. For example, the 2nd row of $H_8$ is:
\begin{equation}
    H_{8,2} = \begin{Bmatrix}
    1 & 0&     1 &  0&     1 &   0 &    1 &   0
    \end{Bmatrix}.
\end{equation}
We break it into a 4 by 8 matrix consisting of 4 sub-rows:
\begin{equation}
    H_{8,2} = \begin{Bmatrix}
1 & 0 & 0 & 0 & 1 & 0 & 0 & 0\\
0 & 0 & 0 & 0 & 0 & 0 & 0 & 0\\
0 & 0 & 1 & 0 & 0 & 0 & 1 & 0\\
0 & 0 & 0 & 0 & 0 & 0 & 0 & 0
\end{Bmatrix}.
\end{equation}
This procedure will quadruple the acquisition time but provide a precise pulse train with less interference between adjacent TFO replicas, and hence a more accurate measurement. One can skip this procedure when a short pulse duration laser is used. The corresponding simulation results can be found in Supplementary Section 4. {As a comparison, if a pulse with a longer pulse duration is used, one has to break each row of Hadamard matrix into more sub-rows, which will lead to an increased data acquisition time}. It is noteworthy that the original Hadamard matrix consists of both 1 and -1 elements. However, since our DMD can only encode non-negative values, we use shifted Hadamard matrix in our temporal encoding: $H_{\text{shift}} = (H+1)/2$. \par

The sampling rate of our system in is 10 Hz (for data in Fig. 4 and 5). Even though the sampling speed is faster than our delay stage ($\sim$2 Hz), we do not attain the potential speed that could be attained with TFO gate. To achieve kHz-level speed, the hardware of our system, including a faster data acquisition system and a more sensitive THz detector, has to be upgraded with details given in Supplementary Section 6.

\section*{Data availability}
The data that support the findings of this study are available from the corresponding author upon reasonable request.

\section*{Code availability}
All relevant computer codes that support the findings of this study are available from the corresponding author upon reasonable request.

\section*{Acknowledgments}
This project is funded by Office of Naval Research (ONR) under grants N00014-17-1-2443, 2204-202-2023940, and N00014-19-1-2247. Jianming Dai at Tianjin University acknowledges support from the National Natural Science Foundation of China under Grant No. 61875151 and the National Key Research and Development Program of China under Grant No. 2017YFA0701000. \par

We acknowledge helpful discussions with Yiyu Zhou and Yiwen E. Jiapeng Zhao thanks Kaia Williams for updating the program used for data acquisition. Robert W. Boyd acknowledges support from the Canada Research Chairs Program and the National Science and Engineering Research Council of Canada. Boris Braverman acknowledges the support of the Banting Postdoctoral Fellowship.

\section*{Author contributions}
J.Z. conceived and developed the concept, carried out the simulation and designed the experiment. J.Z. and J.D. carried out the experiment and performed the measurements. J.D. optimized the system. J.Z. and B.B. analyzed and interpreted the data with the help from R.W.B. and X.Z. R.W.B. and X.Z. supervised the research. All authors contributed to the writing of this manuscript.

\section*{Competing interests}
The authors declare no conflicts of interest.

\bibliography{sample}

\begin{thebibliography}{10}
\newcommand{\enquote}[1]{``#1''}

\bibitem{altmann2018quantum}
Y.~Altmann, S.~McLaughlin, M.~J. Padgett, V.~K. Goyal, A.~O. Hero, and
  D.~Faccio, \enquote{Quantum-inspired computational imaging,}
  {\protect\JournalTitle{Science}} \textbf{361} (2018).

\bibitem{edgar2019principles}
M.~P. Edgar, G.~M. Gibson, and M.~J. Padgett, \enquote{Principles and prospects
  for single-pixel imaging,} {\protect\JournalTitle{Nature Photonics}}
  \textbf{13}, 13--20 (2019).

\bibitem{gibson2020single}
G.~M. Gibson, S.~D. Johnson, and M.~J. Padgett, \enquote{Single-pixel imaging
  12 years on: a review,} {\protect\JournalTitle{Optics Express}} \textbf{28},
  28190--28208 (2020).

\bibitem{candes2008introduction}
E.~J. Cand{\`e}s and M.~B. Wakin, \enquote{An introduction to compressive
  sampling,} {\protect\JournalTitle{IEEE Signal Processing Magazine}}
  \textbf{25}, 21--30 (2008).

\bibitem{wu1995free}
Q.~Wu and X.-C. Zhang, \enquote{Free-space electro-optic sampling of terahertz
  beams,} {\protect\JournalTitle{Applied Physics Letters}} \textbf{67},
  3523--3525 (1995).

\bibitem{teo2015invited}
S.~M. Teo, B.~K. Ofori-Okai, C.~A. Werley, and K.~A. Nelson, \enquote{Invited
  article: Single-shot thz detection techniques optimized for multidimensional
  thz spectroscopy,} {\protect\JournalTitle{Review of Scientific Instruments}}
  \textbf{86}, 051301 (2015).

\bibitem{kane1993characterization}
D.~J. Kane and R.~Trebino, \enquote{Characterization of arbitrary femtosecond
  pulses using frequency-resolved optical gating,} {\protect\JournalTitle{IEEE
  Journal of Quantum Electronics}} \textbf{29}, 571--579 (1993).

\bibitem{iaconis1998spectral}
C.~Iaconis and I.~A. Walmsley, \enquote{Spectral phase interferometry for
  direct electric-field reconstruction of ultrashort optical pulses,}
  {\protect\JournalTitle{Optics Letters}} \textbf{23}, 792--794 (1998).

\bibitem{kauffman1994time}
M.~Kauffman, W.~Banyai, A.~Godil, and D.~Bloom, \enquote{Time-to-frequency
  converter for measuring picosecond optical pulses,}
  {\protect\JournalTitle{Applied Physics Letters}} \textbf{64}, 270--272
  (1994).

\bibitem{kolner1989temporal}
B.~H. Kolner and M.~Nazarathy, \enquote{Temporal imaging with a time lens,}
  {\protect\JournalTitle{Optics Letters}} \textbf{14}, 630--632 (1989).

\bibitem{kolner1994space}
B.~H. Kolner, \enquote{Space-time duality and the theory of temporal imaging,}
  {\protect\JournalTitle{IEEE Journal of Quantum Electronics}} \textbf{30},
  1951--1963 (1994).

\bibitem{bennett1999upconversion}
C.~Bennett and B.~Kolner, \enquote{Upconversion time microscope demonstrating
  103$\times$ magnification of femtosecond waveforms,}
  {\protect\JournalTitle{Optics Letters}} \textbf{24}, 783--785 (1999).

\bibitem{salem2008optical}
R.~Salem, M.~A. Foster, A.~C. Turner, D.~F. Geraghty, M.~Lipson, and A.~L.
  Gaeta, \enquote{Optical time lens based on four-wave mixing on a silicon
  chip,} {\protect\JournalTitle{Optics Letters}} \textbf{33}, 1047--1049
  (2008).

\bibitem{foster2008silicon}
M.~A. Foster, R.~Salem, D.~F. Geraghty, A.~C. Turner-Foster, M.~Lipson, and
  A.~L. Gaeta, \enquote{Silicon-chip-based ultrafast optical oscilloscope,}
  {\protect\JournalTitle{Nature}} \textbf{456}, 81--84 (2008).

\bibitem{foster2009ultrafast}
M.~A. Foster, R.~Salem, Y.~Okawachi, A.~C. Turner-Foster, M.~Lipson, and A.~L.
  Gaeta, \enquote{Ultrafast waveform compression using a time-domain
  telescope,} {\protect\JournalTitle{Nature Photonics}} \textbf{3}, 581--585
  (2009).

\bibitem{okawachi2012asynchronous}
Y.~Okawachi, R.~Salem, A.~R. Johnson, K.~Saha, J.~S. Levy, M.~Lipson, and A.~L.
  Gaeta, \enquote{Asynchronous single-shot characterization of
  high-repetition-rate ultrafast waveforms using a time-lens-based temporal
  magnifier,} {\protect\JournalTitle{Optics Letters}} \textbf{37}, 4892--4894
  (2012).

\bibitem{tikan2018single}
A.~Tikan, S.~Bielawski, C.~Szwaj, S.~Randoux, and P.~Suret,
  \enquote{Single-shot measurement of phase and amplitude by using a heterodyne
  time-lens system and ultrafast digital time-holography,}
  {\protect\JournalTitle{Nature Photonics}} \textbf{12}, 228--234 (2018).

\bibitem{shirai2010temporal}
T.~Shirai, T.~Set{\"a}l{\"a}, and A.~T. Friberg, \enquote{Temporal ghost
  imaging with classical non-stationary pulsed light,}
  {\protect\JournalTitle{Journal of the Optical Society of America B}}
  \textbf{27}, 2549--2555 (2010).

\bibitem{ryczkowski2016ghost}
P.~Ryczkowski, M.~Barbier, A.~T. Friberg, J.~M. Dudley, and G.~Genty,
  \enquote{Ghost imaging in the time domain,} {\protect\JournalTitle{Nature
  Photonics}} \textbf{10}, 167--170 (2016).

\bibitem{devaux2016computational}
F.~Devaux, P.-A. Moreau, S.~Denis, and E.~Lantz, \enquote{Computational
  temporal ghost imaging,} {\protect\JournalTitle{Optica}} \textbf{3}, 698--701
  (2016).

\bibitem{o2017differential}
Y.~O-oka and S.~Fukatsu, \enquote{Differential ghost imaging in time domain,}
  {\protect\JournalTitle{Applied Physics Letters}} \textbf{111}, 061106 (2017).

\bibitem{ryczkowski2017magnified}
P.~Ryczkowski, M.~Barbier, A.~T. Friberg, J.~M. Dudley, and G.~Genty,
  \enquote{Magnified time-domain ghost imaging,} {\protect\JournalTitle{APL
  Photonics}} \textbf{2}, 046102 (2017).

\bibitem{xu2018detecting}
Y.-K. Xu, S.-H. Sun, W.-T. Liu, G.-Z. Tang, J.-Y. Liu, and P.-X. Chen,
  \enquote{Detecting fast signals beyond bandwidth of detectors based on
  computational temporal ghost imaging,} {\protect\JournalTitle{Optics
  Express}} \textbf{26}, 99--107 (2018).

\bibitem{wu2019temporal}
H.~Wu, P.~Ryczkowski, A.~T. Friberg, J.~M. Dudley, and G.~Genty,
  \enquote{Temporal ghost imaging using wavelength conversion and two-color
  detection,} {\protect\JournalTitle{Optica}} \textbf{6}, 902--906 (2019).

\bibitem{tian2020acoustic}
Y.~Tian, H.~Ge, X.-J. Zhang, X.-Y. Xu, M.-H. Lu, Y.~Jing, and Y.-F. Chen,
  \enquote{Acoustic ghost imaging in the time domain,}
  {\protect\JournalTitle{Physical Review Applied}} \textbf{13}, 064044 (2020).

\bibitem{jiang1998single}
Z.~Jiang and X.-C. Zhang, \enquote{Single-shot spatiotemporal terahertz field
  imaging,} {\protect\JournalTitle{Optics Letters}} \textbf{23}, 1114--1116
  (1998).

\bibitem{shan2000single}
J.~Shan, A.~S. Weling, E.~Knoesel, L.~Bartels, M.~Bonn, A.~Nahata, G.~A.
  Reider, and T.~F. Heinz, \enquote{Single-shot measurement of terahertz
  electromagnetic pulses by use of electro-optic sampling,}
  {\protect\JournalTitle{Optics Letters}} \textbf{25}, 426--428 (2000).

\bibitem{kim2006single}
K.~Kim, B.~Yellampalle, G.~Rodriguez, R.~Averitt, A.~Taylor, and J.~Glownia,
  \enquote{Single-shot, interferometric, high-resolution, terahertz field
  diagnostic,} {\protect\JournalTitle{Applied Physics Letters}} \textbf{88},
  041123 (2006).

\bibitem{kim2007single}
K.~Kim, B.~Yellampalle, A.~Taylor, G.~Rodriguez, and J.~Glownia,
  \enquote{Single-shot terahertz pulse characterization via two-dimensional
  electro-optic imaging with dual echelons,} {\protect\JournalTitle{Optics
  Letters}} \textbf{32}, 1968--1970 (2007).

\bibitem{kawada2011single}
Y.~Kawada, T.~Yasuda, A.~Nakanishi, K.~Akiyama, and H.~Takahashi,
  \enquote{Single-shot terahertz spectroscopy using pulse-front tilting of an
  ultra-short probe pulse,} {\protect\JournalTitle{Optics Express}}
  \textbf{19}, 11228--11235 (2011).

\bibitem{zheng2017common}
S.~Zheng, X.~Pan, Y.~Cai, Q.~Lin, Y.~Li, S.~Xu, J.~Li, and D.~Fan,
  \enquote{Common-path spectral interferometry for single-shot terahertz
  electro-optics detection,} {\protect\JournalTitle{Optics Letters}}
  \textbf{42}, 4263--4266 (2017).

\bibitem{takeshi2005asynchronous}
T.~Yasui, E.~Saneyoshi, and T.~Araki, \enquote{Asynchronous optical sampling
  terahertz time-domain spectroscopy for ultrahigh spectral resolution and
  rapid data acquisition,} {\protect\JournalTitle{Applied Physics Letters}}
  \textbf{87}, 061101 (2005).

\bibitem{bartels2007ultrafast}
A.~Bartels, R.~Cerna, C.~Kistner, A.~Thoma, F.~Hudert, C.~Janke, and
  T.~Dekorsy, \enquote{Ultrafast time-domain spectroscopy based on high-speed
  asynchronous optical sampling,} {\protect\JournalTitle{Review of Scientific
  Instruments}} \textbf{78}, 035107 (2007).

\bibitem{dammann1971high}
H.~Dammann and K.~G{\"o}rtler, \enquote{High-efficiency in-line multiple
  imaging by means of multiple phase holograms,} {\protect\JournalTitle{Optics
  Communications}} \textbf{3}, 312--315 (1971).

\bibitem{prongue1992optimized}
D.~Prongu{\'e}, H.-P. Herzig, R.~D{\"a}ndliker, and M.~T. Gale,
  \enquote{Optimized kinoform structures for highly efficient fan-out
  elements,} {\protect\JournalTitle{Applied Optics}} \textbf{31}, 5706--5711
  (1992).

\bibitem{romero2007theory}
L.~A. Romero and F.~M. Dickey, \enquote{Theory of optimal beam splitting by
  phase gratings. i. one-dimensional gratings,} {\protect\JournalTitle{Journal
  of the Optical Society of America A}} \textbf{24}, 2280--2295 (2007).

\bibitem{mirhosseini2013efficient}
M.~Mirhosseini, M.~Malik, Z.~Shi, and R.~W. Boyd, \enquote{Efficient separation
  of the orbital angular momentum eigenstates of light,}
  {\protect\JournalTitle{Nature Communications}} \textbf{4}, 1--6 (2013).

\bibitem{murate2018adaptive}
K.~Murate, M.~J. Roshtkhari, X.~Ropagnol, and F.~Blanchard, \enquote{Adaptive
  spatiotemporal optical pulse front tilt using a digital micromirror device
  and its terahertz application,} {\protect\JournalTitle{Optics Letters}}
  \textbf{43}, 2090--2093 (2018).

\bibitem{stantchev2020real}
R.~I. Stantchev, X.~Yu, T.~Blu, and E.~Pickwell-MacPherson, \enquote{Real-time
  terahertz imaging with a single-pixel detector,}
  {\protect\JournalTitle{Nature Communications}} \textbf{11}, 1--8 (2020).

\bibitem{zhao2019spatial}
J.~Zhao, E.~Yiwen, K.~Williams, X.-C. Zhang, and R.~W. Boyd, \enquote{Spatial
  sampling of terahertz fields with sub-wavelength accuracy via probe-beam
  encoding,} {\protect\JournalTitle{Light: Science \& Applications}}
  \textbf{8}, 1--8 (2019).

\bibitem{sun2017russian}
M.-J. Sun, L.-T. Meng, M.~P. Edgar, M.~J. Padgett, and N.~Radwell, \enquote{A
  russian dolls ordering of the hadamard basis for compressive single-pixel
  imaging,} {\protect\JournalTitle{Scientific Reports}} \textbf{7}, 1--7
  (2017).

\end{thebibliography}

\end{document}